\documentclass[useAMS]{mn2e}
\usepackage{amsmath}
\usepackage{textcomp}
\usepackage{amssymb}
\usepackage[pdftex]{graphicx}
\usepackage{amssymb}

\usepackage[margin=.8in, paperwidth=8.5in, paperheight=11in]{geometry}
\usepackage{multicol}
\usepackage{multirow}
\usepackage{epsfig}
\usepackage{graphicx}
\usepackage{dblfloatfix}
\usepackage{accents}
\usepackage{bigdelim}
\newcommand*{\dt}[1]{\accentset{\mbox{\Large\bfseries .}}{#1}}

\begin{document}

\title{Counter-Rotating Accretion Discs}
\author[S. Dyda et al.]
{\parbox{\textwidth}{S.~Dyda,$^{1}$
R.V.E.~Lovelace,$^{2}$
G.V.~Ustyugova,$^{3}$
M.M.~Romanova,$^{2}$\\
 \& A.V.~Koldoba$^{3}$}\vspace{0.3cm}\\
 $^{1}$Department of Physics, Cornell University, Ithaca, NY 14853:email: sd449@cornell.edu\\
 $^{2}$Department of Astronomy, Cornell University, Ithaca, NY 14853\\
$^{3}$Keldysh Institute for Applied Mathematics, Moscow, Russia\\
}

\date{\today}
\pagerange{\pageref{firstpage}--\pageref{lastpage}}
\pubyear{2014}

\label{firstpage}

\maketitle

\begin{abstract}

    Counter-rotating discs can arise from the accretion of a counter-rotating
 gas cloud onto the surface of an existing co-rotating disc or from the
 counter-rotating gas moving radially inward to the outer edge of an existing disc.  
      At the interface,  the two components mix to produce gas or
plasma with zero net angular momentum which tends to free-fall
towards the disc center.   We  discuss high-resolution axisymmetric
hydrodynamic simulations of viscous counter-rotating disc for  cases
where the two components are vertically separated and radially 
separated.  The viscosity is described by an isotropic $\alpha-$viscosity
including all terms in the viscous stress tensor.   For the
vertically separated components a shear layer forms between them
the middle of this layer in free-fall to the disc center.  The
accretion rates are increased by factors $\sim 10^2-10^4$ over that
for a conventional disc rotating in one direction with the same
viscosity.
The vertical width  of the shear layer and the accretion rate are strongly 
dependent on the viscosity and the mass fraction of the counter-rotating gas.
    In the case of radially separated components where the inner disc
co-rotates and the outer disc rotates in the opposite direction,
a gap between the two components opens and closes quasi-periodically.
  The accretion rates are $\gtrsim 25$ times larger than those for a disc
rotating in one direction with the same viscosity.

\end{abstract}

\begin{keywords} accretion discs-galaxies: evolution-galaxies:formation-galaxies:nuclei-galaxies:spiral
\end{keywords}

\section{Introduction}

Commonly considered models of accretion discs have gas
rotating in one direction with a turbulent viscosity acting to
transport angular momentum outward and matter inward.
   However, observations
indicate that there are  more complicated, possibly
transient,  disc structures
 on a galactic scale.
        Observations of
normal galaxies have revealed co/counter-rotating gas/gas,
 gas/stars, and stars/stars discs
   in many galaxies of all morphological types ---
ellipticals, spirals, and irregulars (Rubin 1994a, 1994b; Galletta
1996;  review by Corsini 2014).  
  There is limited evidence  of  counter-rotating gas components in discs around stars (Remijan and Hollis 2006; Tang et al. 2012).
     Theory and simulations of
 co/counter-rotating gas/gas discs predict
enormously enhanced accretion rates resulting from the
presence of the two components which can give rise to
outbursts from these discs (Lovelace and Chou 1996; Kuznetsov et
al. 1999; Nixon, King, and Price 2012; Vorobyov, Lin, and Guedel 2014).

   Counter-rotation in disc galaxies appears in a variety
of configurations including cases where (1) two cospatial 
populations of stars rotate in opposite directions,
(2) the stars rotate in one direction and the gas rotates
in the opposite direction, and (3) two spatially separated
gas discs rotate in opposite directions (Corsini 2014).
      An example of case (1) is the galaxy
NGC 4550  which has  cospatial counter-rotating stellar discs (Rubin, Graham, and Kenney 1992).
    Examples of case (2) include the galaxy NGC 3626
(Ciri, Bettoni, and Galetta 1995) and NGC 4546 (Sage and Galletta 1994).
     An example of case (3) is the ``Evil Eye'' galaxy
NGC 4826 in which the direction of rotation of the gas
reverses going from the inner ($180$ km s$^{-1}$) to the outer disc
($-200$ km s$^{-1}$) with an inward radial accretion speed of
more than $100$ km s$^{-1}$in the transition zone, whereas the
stars at all radii rotate in the same direction as the gas in the
inner disc, which has a radius of $\sim 1200$ pc (Braun et al. 1994;
Rubin 1994b).   
     
     On a stellar mass scale, Remijan and Hollis (2006) and  Tang et al. (2012)  found evidence of  accretion of counter-rotating gas in star forming systems.   
    Accreted counter-rotating matter may
encounter an existing co-rotating disc, for example, in low-mass X-ray
binary sources where the accreting, magnetized rotating
neutron stars are observed to jump between states where
they spin up and those where they spin down. Nelson et al.
(1997) and Chakrabarty et al. (1997) have proposed
that the change from spin-up to spin-down results from a
reversal of the angular momentum of the wind-supplied
accreting matter.

       In the formation of massive black holes in the nuclei
 of active galaxies, King and Pringle (2006, 2007) argue
 that in order for the rapid growth of the massive black holes
 (observed at high red-shifts) to be compatible with the
 Eddington limit on their emission, the gas accretion is from
 a sequence clouds with randomly oriented angular momenta.
 This process would lead to the formation of a disc with the
 rotation direction alternating as a function of radius. 

    A number of theoretical and computer simulation studies have
 explored the different physical processes arising in counter rotating
 discs.  In the case of counter rotating cospatial stellar discs, the two-stream instability is predicted to couple the two components and give rise to low azimuthal mode number ($m=1,~2$) spiral waves (Lovelace, Jore, \& Haynes 1997).  
 $N-$body simulations of counter rotating stellar
discs show a prominent $m=1$ spiral wave (Comins et al. 1997). 
   The two-stream instability is also important for cospatial stellar/gas
 discs (Lovelace et al. 1997).
 
     In the case of counter-rotating gas discs, the two components are necessarily
 spatially separated because otherwise the ensuing strong shocks would
 destroy the discs.  The counter-rotating gas may come from later infall of 
 ``new''  gas onto a preexisting co-rotating disc.
     A self-similar analytic solution was
derived  for accretion discs with isotropic
$\alpha-$viscosity (Shakura and Sunyaev 1973)  where the top
half of the disc ($Z>0$) rotates in one direction and the bottom half rotates in the other direction  (Lovelace and Chou 1996).  The rotation can be Keplerian for accretion
to a star or black hole or it can be a flat rotation for the case of a disc galaxy.
    The gas in the super-sonic shear layer between the two components 
has  nearly free-fall inward radial velocity.    The accretion rate is enhanced relative to a conventional disc by a factor of the order of  $(R/h)^2 \alpha^{-1/2} \gg 1$,
where $R$ is the radius and $h$ is the disc's half-thickness.
    The Kelvin-Helmholtz  instability is expected to be important in
such shear layers.   It  has been studied earlier in planar supersonic shear layers (Ray 1981, 1982; Choudhury and Lovelace 1983).  
    Gulati, Saini, and Sridhar (2012) studied the $m=1,~ 2$ modes component Keplerian discs and found them to be unstable and a possible source of lopsided brightness distributions in galaxies. 
        Using smooth particle hydrodynamic simulations,
 Nixon et al. (2012) modeled counter-rotating inner and outer AGN discs that were tilted with respect to one another and found that accretion rates could be $\gtrsim 100$ times greater than if they were planar.
     Also, using smooth particle hydrodynamic simulations,
Alig et al. (2013) modeled counter-rotating gas discs with the aim of
explaining the  spiral-like filaments of gas feeding the galactic center's
black hole  SgrA*.

     Spherical grid,  axisymmetric  hydrodynamic simulations of 
different configurations of counter-rotating discs were carried
out by Kuznetsov et al. (1999).   The interaction of the co- and
counter-rotating components was mediated by the numerical 
viscosity.       The present work discusses high-resolution
axisymmetric hydrodynamic simulations of counter rotating discs
including all components of the viscous stress
tensor for an isotropic $\alpha-$viscosity.

   Section 2 of the paper discusses the numerical methods and the simulation parameters. 
     Section 3 discusses the simulation results for the cases of vertically
and radially separated components.  Section 4 gives the  conclusions.

\section{Model}

\subsection{Basic Equations}

    The  flows are assumed to be described by the
equations of non-relativistic hydrodynamics (HD). 
In a non-rotating reference frame the equations are
\begin{subequations}
 \begin{equation}
\frac{\partial \rho}{\partial t} + \nabla \cdot \left( \rho \mathbf{v} \right) = 0~, 
\end{equation}
\begin{equation}
 \frac{\partial \rho \mathbf{v}}{\partial t} + \nabla \cdot \mathcal{T} = \rho \mathbf{g}~,
\end{equation}
\begin{equation}
\frac{\partial \left( \rho S \right)}{\partial t} + \nabla \cdot \left( \rho  \mathbf{v}S \right) = \mathcal{Q}~.
\end{equation}
\end{subequations}
Here, $\rho$ is the mass density, $S$ is the specific entropy, $\mathbf{v}$ is the flow velocity, $\mathcal{T}$ is the momentum flux density tensor and $\mathcal{Q}$ is the rate of change of entropy per unit volume due to viscous
and Ohmic heating in the disc. 
    We assume that the heating is offset by radiative cooling so that
$Q=0$.
   Also, $\mathbf{g} = -\left(GM/r^2 \right)\hat{\bf r}$
is the gravitational acceleration due to the
central mass $M$.
We model the fluid as a non-relativistic ideal gas with equation of state
\begin{equation}
 S = \ln\left( \frac{p}{\rho^{\gamma}}\right)~,
\end{equation}
where $p$ is the pressure and $\gamma = 5/3$. We use cylindrical coordinates $(R,\phi,Z)$ as these are the coordinates used by our code. 

The viscosity of the gas is considered to be due to turbulent fluctuations of the velocity.  Outside of the disc, the gas is assumed to be ideal with negligible viscosity.
The turbulent coefficient is parameterized using the  $\alpha$-model of Shakura and Sunyaev (1973). The turbulent kinematic viscosity is 
\begin{equation}
\nu_t = \alpha_{\nu} \frac{c_s^2}{\Omega_K}~,
\end{equation}
where $c_s$ is the midplane  sound speed, $\Omega_K$ is the Keplerian angular velocity at the given radii and $\alpha_{\nu}\leq 1$ is a dimensionless constant.

The momentum flux density tensor is given by
\begin{equation}
 \mathcal{T}_{ik} =p\delta_{ik}+ \rho v_i v_k + \tau_{ik}~,
\end{equation}
where $\tau_{ik}$ is the viscous stress contribution from the turbulent fluctuations of the velocity. 
    As mentioned, we assume that these can be represented in the same
way as the collisional viscosity by substitution of the turbulent viscosity. The leading order contributions to the viscous stress arise from large velocity gradients. In a Keplerian type disc these will be dominated by the azimuthal terms $v_{\phi} \sim v_{K}$. However, in a counter-rotating disc the boundary layer between the counter-rotating components experience nearly free fall velocity. Therefore we include all viscous terms involving $v_R$ and $v_{\phi}$
   The leading order contribution to the momentum flux density from 
turbulence are therefore
\begin{align}
\tau_{R\phi} = -\nu_t \rho R \frac{\partial \Omega}{\partial R} ~, \hspace{1.2cm}
\tau_{Z \phi} = - \nu_t \rho R \frac{\partial \Omega}{\partial z}~, \nonumber \\
 \tau_{Z R} = - \nu_t \rho \frac{\partial v_R}{\partial Z}~, \hspace{1.2cm}
\tau_{R R} = - 2 \nu_t \rho \frac{\partial v_R}{\partial R}~, \nonumber \\
 \tau_{\phi \phi} = - 2 \nu_t \rho \frac{v_R}{R} ~, \hspace{4cm}
\end{align}
where $\Omega = v_{\phi}/R$ is the angular velocity of the gas.

    The transition from the viscous disc to the non-viscous
corona is handled  by multiplying the viscosity  by a dimensionless factor $\xi(\rho)$ which varies
smoothly from $\xi=1$ for $\rho \geq \rho_d=0.75\rho(R,Z=0)$   to
$\xi =0$ for $\rho\leq 0.25 \rho_d$ as described in Appendix B
of Lii, Romanova, \& Lovelace (2012).

\begin{figure}
                \centering
                \includegraphics[width=0.5\textwidth]{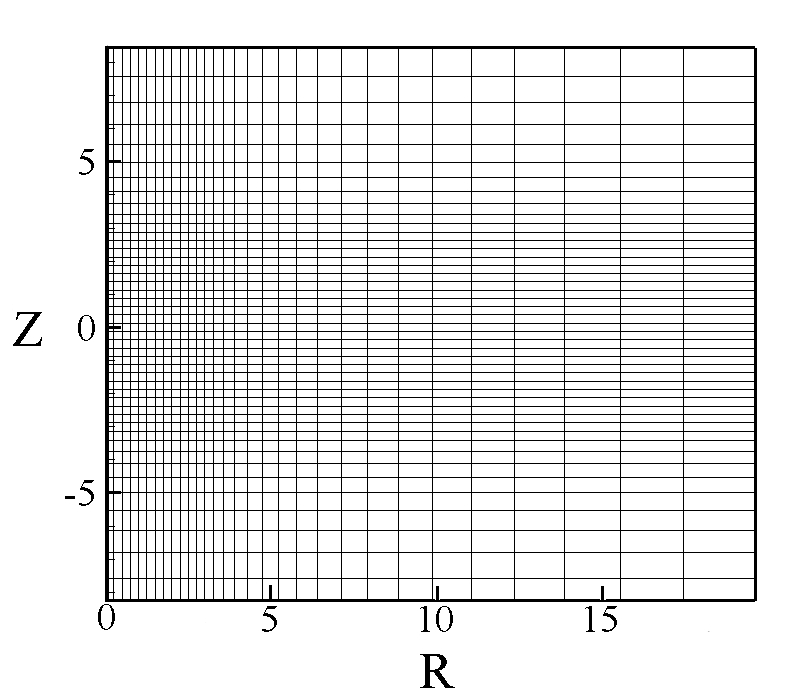}
        \caption{ Sparse version of our grid showing $20\%$ 
of the cells used.    The simulations used $N_R\times N_Z =150 \times 250$ grid cells.}
\label{fig:mesh}
\end{figure}

\subsection{Initial Conditions}

\subsubsection{Matter Distribution}
Initially the matter of the disc and corona are assumed to be in mechanical equilibrium (Romanova et al. 2002). 
The initial density distribution is taken to be barotropic with
\begin{equation}
  \rho(p) =
  \begin{cases}
   p/T_{\rm{disc}} & p>p_b ~~{ \rm and} ~~ R \geq R_d~, \\
   p/T_{\rm{cor}} & p<p_b ~~ {\rm or} ~~ R  \leq R_d~,
  \end{cases}
\end{equation}
where $p_b$ is the level surface of pressure that separates the cold matter of the disc from the hot matter of the corona and $R_d$ is the inital inner disc radius. At this surface
the density has an initial step discontinuity from value $p/T_{\rm{disc}}$ to $p/T_{\rm{cor}}$.

Because the density distribution is barotropic, the initial angular velocity magnitude is a constant on coaxial cylindrical surfaces about the $z-$axis. Consequently, the pressure 
can  be determined from the Bernoulli  equation
\begin{equation}
 F(p) + \Phi + \Phi_c = \rm{const}~,
\end{equation}
where $\Phi = -GM/\sqrt{R^2 + Z^2}$ is the gravitational potential, $\Phi_c = \int_{R}^{\infty}\xi d\xi ~\omega^2(\xi)$ is the 
centrifugal potential, and
\begin{equation}
  F(p) =
  \begin{cases}
   T_{\rm{disc}}\ln(p/p_b) & p>p_b ~~ {\rm and} ~~ R \geq R_d~, \\
   T_{\rm{cor}}\ln(p/p_b) & p<p_b ~~{\rm or} ~~ R \leq R_d~.
  \end{cases}
\end{equation}
    The initial half-thickness of the disc $h$ is taken to
be  $h/R = 0.200$, and the inner disc radius $R_d = 2$.  
   We have also carried out simulations for $h/R=0.1$ and
 $R_d =2$ for the vertically and radially separated components
 and conclude that the results present below do not depend
 significantly on $h/R$ in this range.

\subsubsection{Angular Velocity}

The magnitude of the initial angular velocity is constant along cylinders of constant $R$.
Inside of $R_d$, the matter rotates rigidly with angular velocity of the star
\begin{equation}
 \Omega = (1-0.003)\Omega_K(R_d) \hspace{1cm} R  \leq R_d~.
\end{equation}
Inside the disc the angular velocity of the disc is slightly sub-Keplerian. 

    For the simulations of vertically separated co- and counter-rotating
components,  we consider a positive angular velocity of
the lower part of the disc.  The upper counter-rotating part
of the disc has a negative angular velocity.    That is, 
\begin{equation}
 \Omega = 
\begin{cases}
- (1-0.003)\Omega_K(R) \hspace{1cm} Z>0 ~ {\rm and} ~ \rho \leq \rho_c~, \\
\hspace{.25cm}(1-0.003)\Omega_K(R) \hspace{1cm} \rm{elsewhere}~.
\end{cases}
\end{equation} 
We parametrize these runs in terms of the inital ratio of disc mass counter-rotating
with the star to the disc mass rotating with the star which we define as $\Delta$. The counter-rotating floor density 
$\rho_c$ is determined by the choice of $\Delta$.

    For the simulations where the co- and counter-rotating components
are radially separated at the radius $R_c$, we  consider
\begin{equation}
 \Omega = 
\begin{cases}
\hspace{.25cm} (1-0.003)\Omega_K(R) \hspace{1cm} R_d<R<R_c~, \\
-(1-0.003)\Omega_K(R) \hspace{1.9cm} R>R_c~. ~
\end{cases}
\end{equation}

\begin{figure*}
                \centering
                \includegraphics[width=0.95\textwidth]{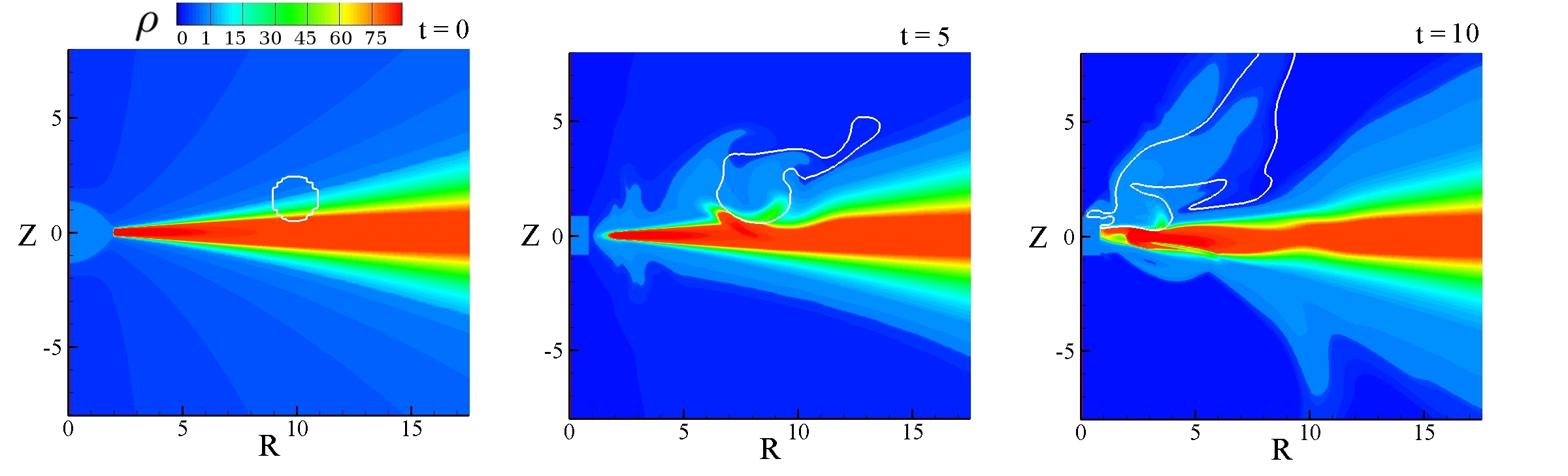}
        \caption{Density $\rho$ (color) and angular velocity
   $\Omega = 0$ countours (white lines) for $t = 0,~5,~10$ for the torus problem.    The initial counter rotating torus (left-hand panel)
  is observed to free-fall into the star (right-hand panel). 
}
\label{fig2}
\end{figure*} 

\subsection{Computational Domain and Boundaries}

Our simulation region has three boundaries: the axis, the surface of the star and the external boundaries. For each dynamical variable we impose a boundary
condition consistent with our physical assumptions.

We assume axisymmetry.
 On the star and the external boundaries we impose free boundary conditions $\partial \mathcal{F}/\partial n = 0$ where $\mathcal{F}$ is a dynamical variable and $n$ is the vector normal to the boundary. 
 
 The hydrodynamic equations are written in dimensionless form so
that the simulation results can be applied to different systems.
    The mass of the central star is taken as the reference unit of mass, $M_0 = M_?$ . The reference length,
$r_0$ , is taken to be the radius of the star. The initial in-
ner radius of the disc is $R_d = 2r_0$ . The reference value
for the velocity is the Keplerian velocity at the radius $r_0$ ,
$v_0 = (GM_0 /r_0 )^{1/2}$ . The reference time-scale is the period of rotation at $r_0$ , $P_0 = 2\pi r_0 /v_0$.

At the external boundary along the edge of the disc $-5 < Z < 5$, we allow new matter to flow into the simulation region. We impose the condition that the matter
must be accreting $v_R < 0$. In the coronal region, we prescribe outflow conditions and allow matter and entropy to exit the simulation region.

Our simulations use a grid $N_R\times N_Z =150 \times 250$ cells. The star has a radius of $1$ in our dimensionless units and is cylindrical in shape. 
     It extends $10$ units above and below the equatorial plane. 
     In the $R-$direction, 
the first $60$ grid cells have length $dR = 0.05$. 
     At larger radii the cell lengths are given recursively by $dR_{i+1} = 1.025 dR_{i}$. Similarly, in the $Z-$direction
the first 30 grid cells above and below the equatorial plane have length $dZ = 0.05$. 
    At larger $Z$, the  cell lengths are given recursively by $dZ_{j+1} = 1.025 dZ_{j}$.    Figure 1 shows
a sparse version of our grid with only every $5$th cell shown.

\begin{figure}
                \centering
                \includegraphics[width=0.45\textwidth]{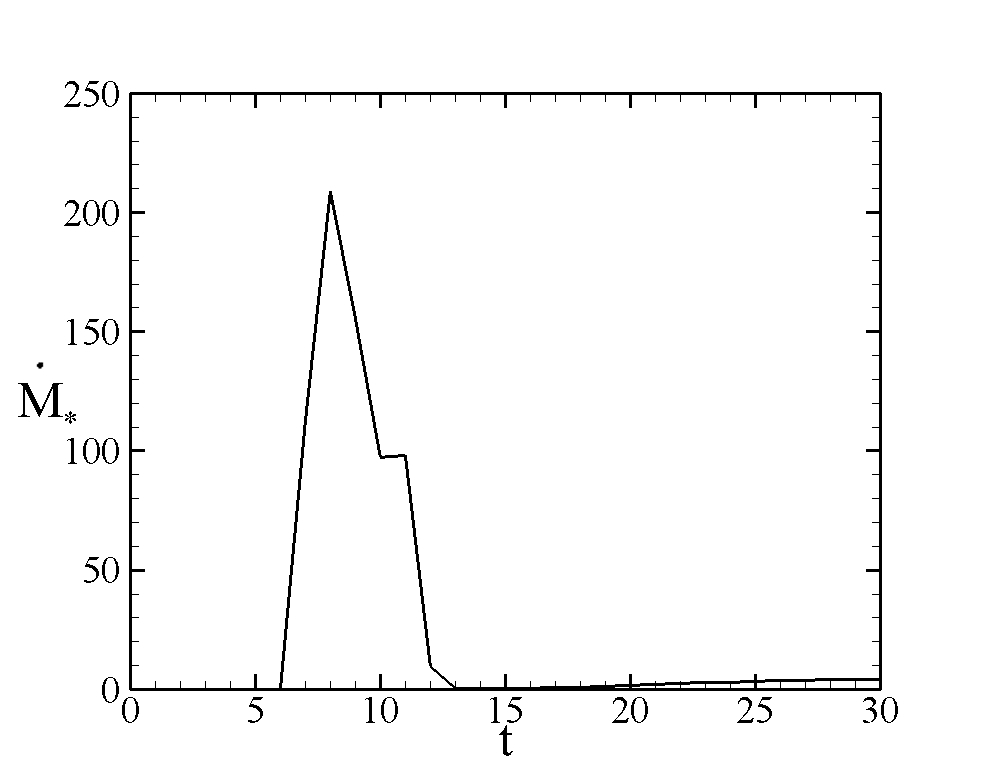}
        \caption{Mass accretion rate to the star $\dot{M}_*$  as a function of time for a counter-rotating torus of matter initially
located at $R_0 = 10$.   After the burst of accretion, accretion to the star
continues at a small rate due to the disc's viscosity.
}
\label{fig:mdotblob}
\end{figure} 

\begin{figure}
                \centering
                \includegraphics[width=0.35\textwidth]{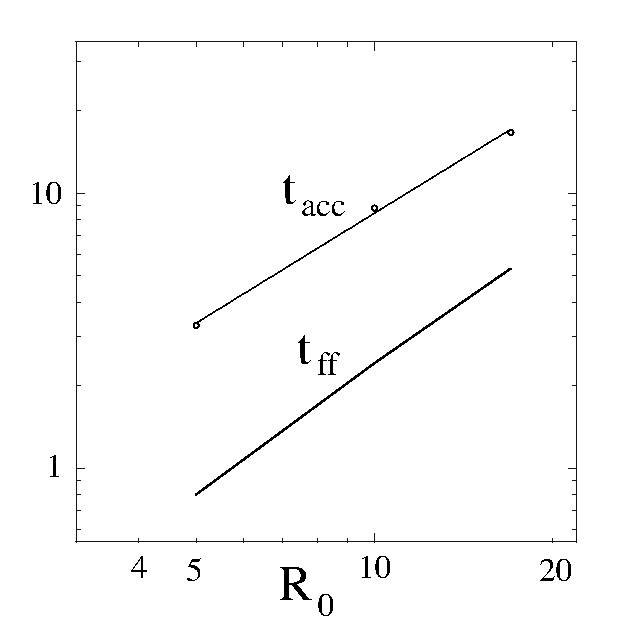}
        \caption{Accretion time $t_{acc}$ (arbitrary scale)  for torus of counter-rotating matter as a function of its initial radius $R_0 = 5,10,17$ and free fall times from $R_0$,  $t_{ff} = [2R_0^3/(9GM)]^{1/2}$. 
}
\label{fig:freefall}
\end{figure} 

\section{Results}

\subsection{Vertically Offset Torus}

    Here, we consider an equilibrium consisting of a co-rotating disc 
with a counter-rotating torus of a small amount of mass
with major axis $R_0$ and minor axis $r \ll R_0$ located at $(R_0,Z_0)$.
    The mass of the torus is $1\%$ of the disc's total mass, and it
is in hydrostatic equilibrium with the disc in the $Z-$direction. 
   This is for the purpose of modeling the accretion of a counter-rotating
cloud onto the surface of a preexisting co-rotating disc.   
   Figure 2 shows the infall of the torus at several times.
    The   torus falls inward over the top surface of the
disc towards the star.    Even though the torus has a
very small mass fraction it induces axisymmetric warping
of the disc.
         When it reaches the star it gives a  large
spike in the stellar accretion rate $\dt{M}_*$ as shown in Fig. 3.

      The infall time of the torus from $R_0$ to the star,
 $t_{\rm acc}(R_0)$,  can be compared to the time of free-fall $t_{\rm ff} \approx [2R_0^3/(9GM_*)]^{1/2}$ for $R_0 \gg R_*$ for a blob
 of non-rotating  matter to free-fall to the star.
 Figure 4 shows the two time scales as a function of $R_0$.
     The accretion time-scale $t_{\rm acc}$ is about a factor of $3$
longer than the free-fall time and it is $ \propto R_0^{1.32}$.
    This makes sense because the free-fall time assumes that the 
torus is initially at rest, whereas in the simulations the torus
is initially counter-rotating.
    The torus first loses  it's angular momentum to
the main disc on a viscous time scale which can be estimated
as $t_\nu \sim t_{\rm ff} (Z/h)^2 \alpha_\nu^{-1}$.  
     This can account
roughly for the factor of 3.  Note that at the interface between
the oppositely rotating components there can be a strong
supersonic Kelvin-Helmholtz instability (Quach, Dyda, \& 
Lovelace 2014) which locally increases the turbulent
viscosity coefficient by a large factor.

\label{sec:wedge}

\begin{table*}
\begin{center}
  \begin{tabular}{ | c  c  c  c  c  c|  |c  c  c  c  c  c  c  c|}
                                                                 \\\hline
$\Delta$&  $\alpha_{\nu}$ & $(\dt{M}_{*})_p$ & $(\dt{m})_p$ & $M_T$ & $t_w$ &&\multicolumn{6}{c}{$\Delta Z(R)$} \\ 
        &                 &              &          &       &       &$t=$&\multicolumn{3}{c}{5} &\multicolumn{3}{|c|}{15}\\
        &                 &              &          &       &       &$R=$& 5 & 10 & 15 & 5 & 10 & 15    \\ \hline \hline
  0.01  &  0.03           & 236          & 23,600   & 5,000   & 50.0  && 0.45    & 0.56  & 0.56    & 0.77  & 1.15       & 1.38        \\
        &  0.1            & 144          &    758   & 2,030  & 38.6  && 0.85    & 1.06   & 1.36    & 1.56  & 2.18 & 2.11        \\ 
        &  0.2            & 68.4         &    79.5  & 1,040  & 45.2  && 1.02    & 1.29   & 1.85    & 2.15  & 2.67       & 2.43           \\
        &  0.3            & 26.1         &    13.4  &  460  & 39.3  && 1.12    & 1.46   & 1.69    & 2.30  & 3.06       & 2.46          \\\hline
  0.02* &  0.1            & 288          &  1,516   &  4,400 & 40.0  && 0.77    & 0.98   & 1.14    & 1.55  & 2.19       & 2.11          \\ \hline 
  0.1   &  0.1            & 1,406         &  7,400   & 31,300& 49.0  && 0.58    & 0.70   & 0.85    & 0.78  & 1.46       & 1.86           \\\hline
  0.5   &  0.03           &  -           &    -     &602,700&   -   && 0.19    & 0.31  & 0.41    & 0.28  & 0.59       & 0.83          \\
        &  0.1            &  -           &    -     &260,000&   -   && 0.37    & 0.60  & 0.75    & 0.50  & 1.17       & 1.65          \\\
        &  0.2            &  -           &    -     & 32,590&   -   && 0.60    & 0.91  & 1.14    & 0.86  & 1.92       & 2.56          \\
        &  0.3            &  -           &    -     & 15,560&   -   && 0.84    & 1.19  & 1.47    & 1.24  & 2.66       & 3.42          \\           
\hline \hline
  \end{tabular}
\end{center} 
\caption{Summary of results for vertically separated 
components for different counter-rotating mass fractions $\Delta$ and viscosity coefficients $\alpha_{\nu}$.    The asterisk  indicates symmetric wedges above and below the midplane. We measure the peak mass accretion rate $(\dot{M}_{*})_p$, peak mass accretion rate normalized to a Keplerian disc $(\dot{m})_p$, the total mass accreted while there is a counter-rotating component $M_T$ and the time required for all counter-rotating matter to accrete $t_w$. For times $t = 5$ and $t = 15$ we measure the thickness of the shear layer $\Delta Z(R)$, at radii $R = 5, 10, 15$. }
\label{table:wedge} 
\end{table*}

\begin{figure*}
                \centering
                \includegraphics[width=0.8\textwidth]{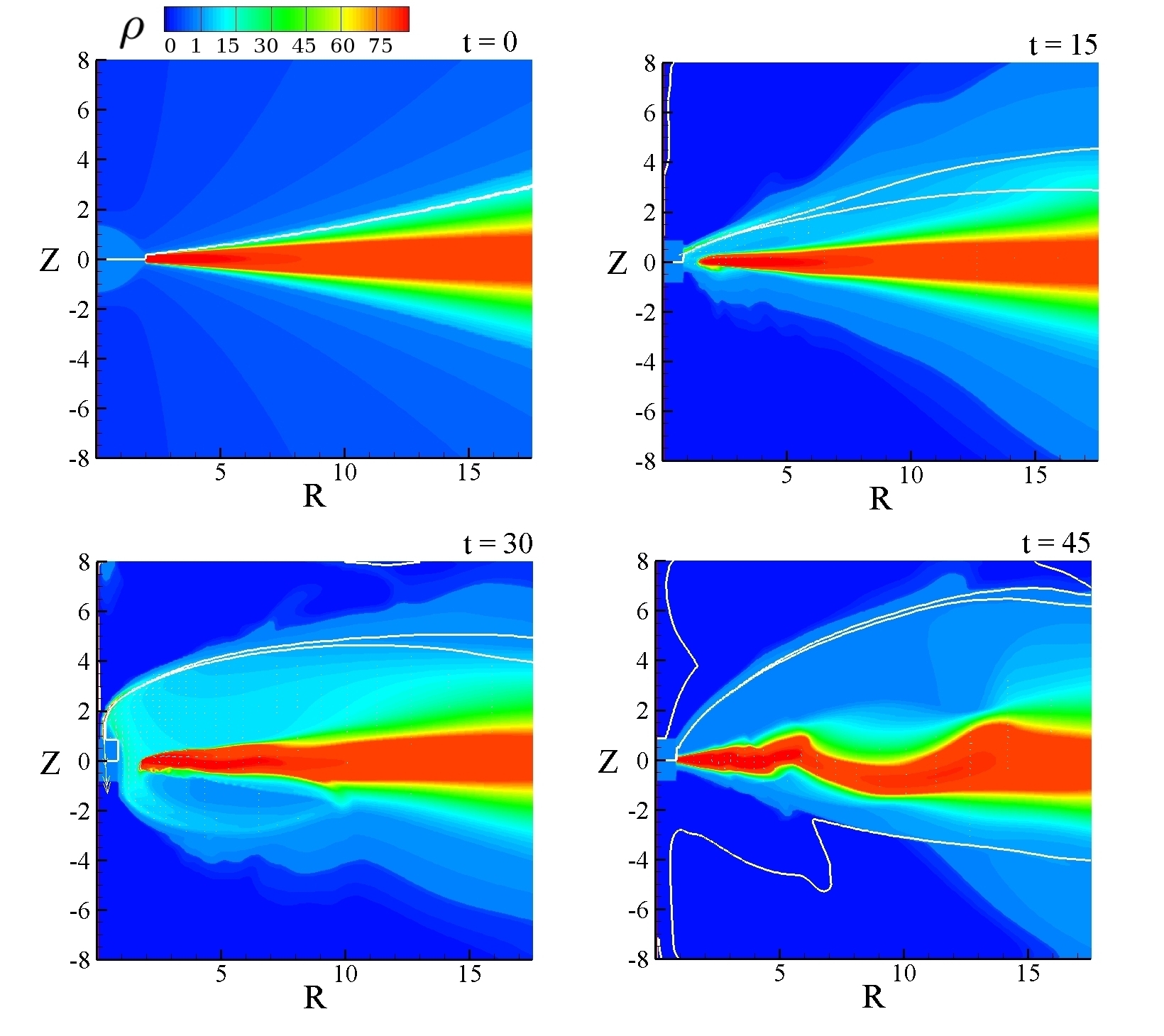}
        \caption{Density $\rho$ (color) and angular velocity $\Omega = \pm 0.01$ countours (white lines) for $t = 0,15,30,45$ for the case $\Delta = 0.01$ and $\alpha_{\nu} = 0.1$.   The initial counter rotating layer in the upper half plane (top left-hand panel), the shear layer accretes to the star (top right-hand panel), moving upwards  (bottom left-hand
 paneel) before all counter-rotating matter has lost its angular momentum to the main disc and oscillations are induced in the disc (bottom right-hand
 panel).   Notice the  large amplitude axisymmetric warping induced
 in the disc. }
\label{fig:wedge_summary}
\end{figure*}
\begin{figure}
                \centering
                \includegraphics[width=.4\textwidth]{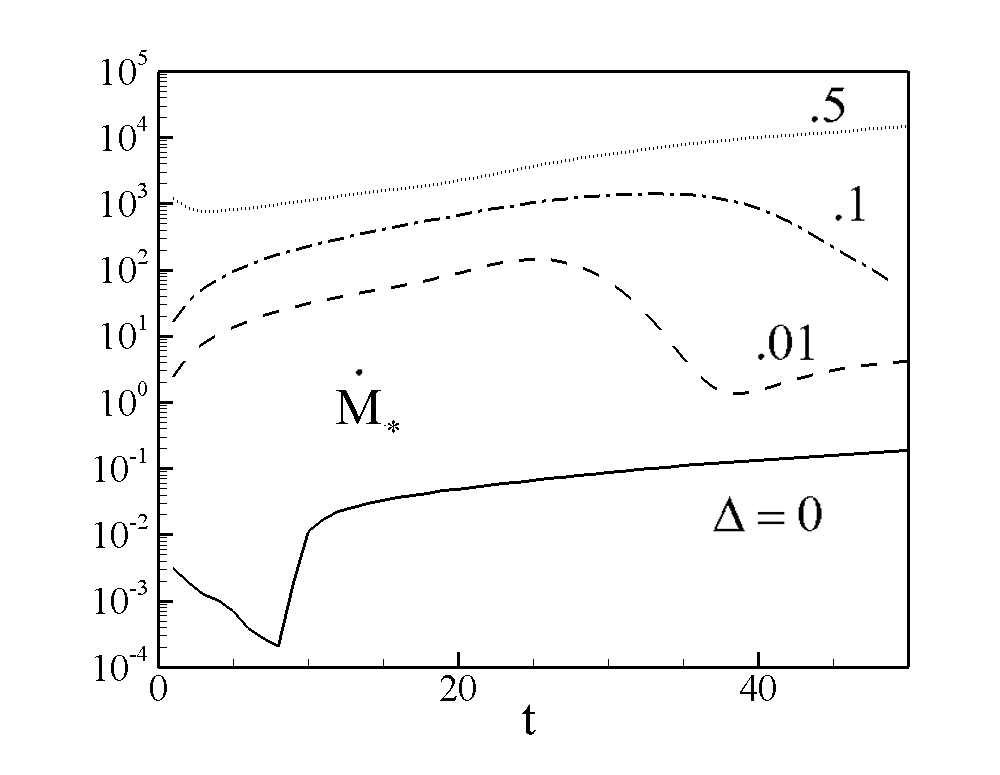}
        \caption{Mass accretion rate $\dot{M}$ as a function of time $t$ for mass fractions $\Delta = 0, ~0.01,~ 0.1,~ 0.5$ and viscosity
 coefficient $\alpha_{\nu} = 0.1.$ 
}
\label{fig:mdot_alpha}
\end{figure}

\begin{figure}
                \centering
                \includegraphics[width=.4\textwidth]{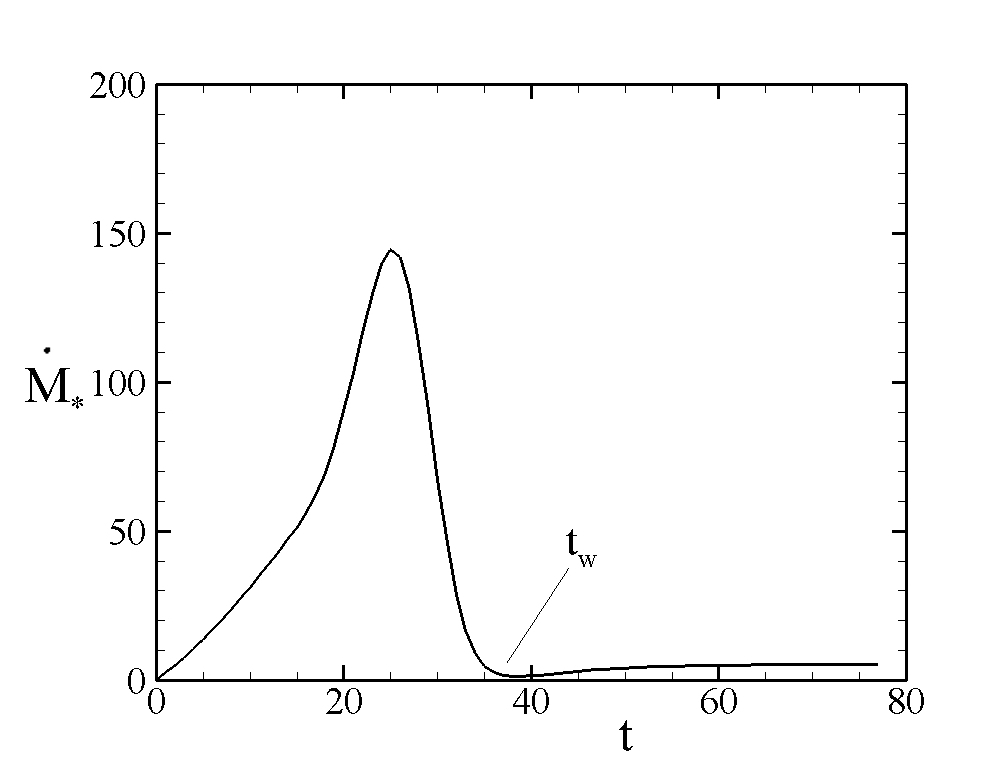}
        \caption{Mass accretion rate to the star $\dot{M}_*$ as a function of time for $\Delta = 0.01$ and $\alpha_{\nu} = 0.1$. 
Here,  $t_w$ indicates the time at which all counter-rotating matter has accreted onto the star.
}
\label{fig:mdot_t}
\end{figure}

\begin{figure}
                \centering
                \includegraphics[width=0.4\textwidth]{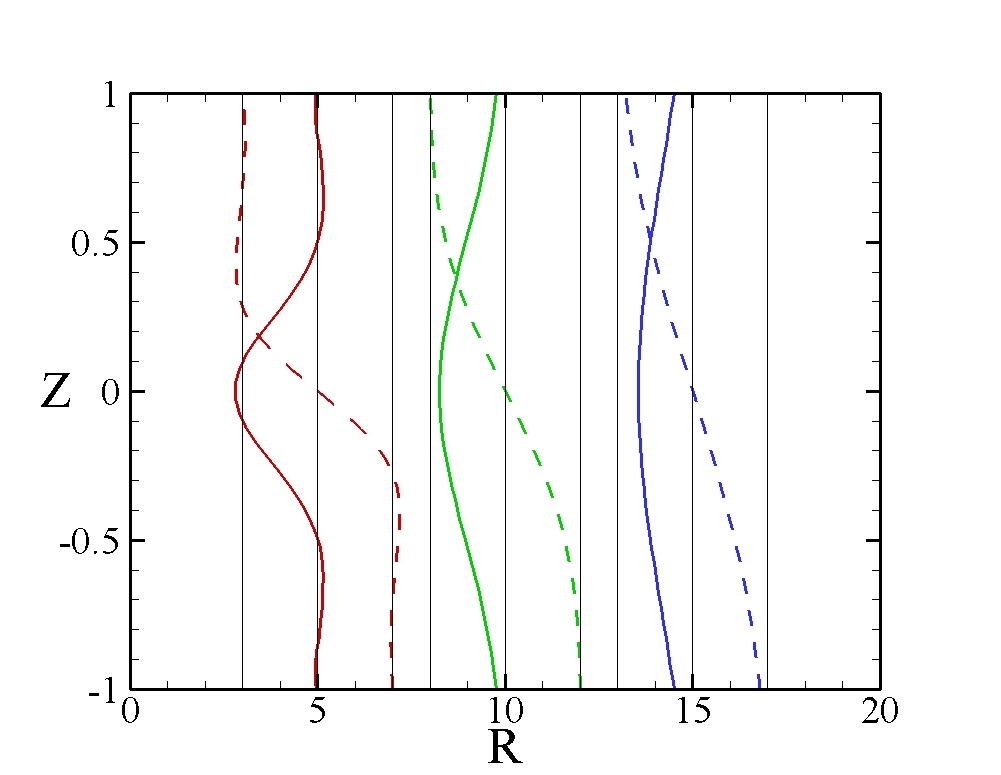}
        \caption{Radial and angular velocity profiles as a function of
$Z$ $v_{r}(R,Z)$ (solid curves) and   $\Omega(R,Z)$ (dashed curves) at radii $R = 5$  (red), $10$ (green) and $15$ (blue) in units of the local Keplerian velocity and angular velocity (indicated
by the vertical lines) for the case $\Delta = 0.5$, $\alpha_{\nu} = 0.1$ at $t = 15$.  The velocity profiles 
are approximately Gaussian and peaked at $Z=0$ where $\Omega=0$.}
\label{fig:v}
\end{figure}

\begin{figure}
                \centering
                \includegraphics[width=.4\textwidth]{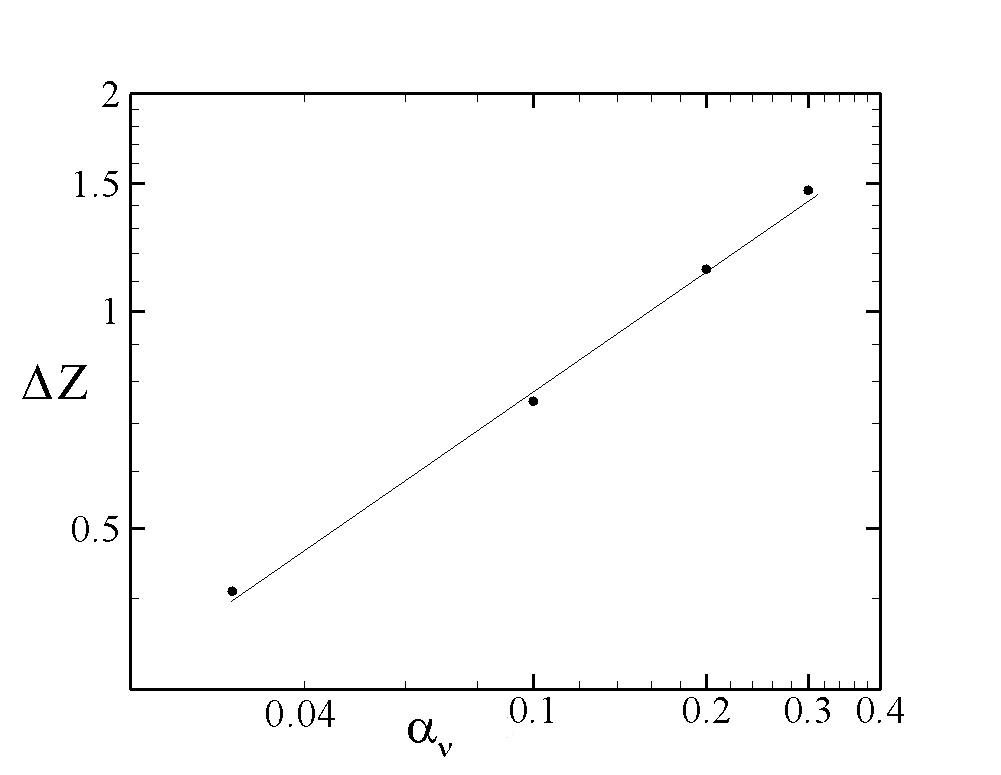}
        \caption{Full width at half-maximum of the  radial velocity destribution $\Delta Z(R)$ as a function of viscosity $\alpha_{\nu}$ for $R = 15$ and $t=15$ for the case $\Delta = 0.5$. The  fit for radii $R = 5,10,15$ and $t = 5, 15$ gives a power law dependence
 $\Delta Z \propto \alpha_{\nu} ^ {0.59 \pm 0.03}$, in general agreement with the self-similar solution $\Delta Z \propto \sqrt{ \alpha_{\nu}}$ of Lovelace and Chou 1996).
}
\label{fig:width}
\end{figure}

\subsection{Vertically Separated Components}

  Here, we consider cases where the co-rotating and counter-rotating
components are vertically separated.   This situation could 
arise from the infall of counter-rotating gas onto the surface
of a preexisting co-rotating disc.
    That is, the disc matter at some distance above 
the midplane  is counter-rotating with
Keplerian velocity while the matter below this height is
co-rotating.
    The fraction of the surface mass density which is counter-rotating
 is denoted by $\Delta$.     A conventional disc rotating in one direction
has  $\Delta = 0$ and a Shakura-Sunyaev accretion rate $\dt{M}_{SS}$
which is a function of the viscosity coefficient $\alpha_\nu$.    
For $\Delta>0$ the mass accretion rate is normalized by the
value  for $\Delta=0$.  That is, $\dt{m}\equiv \dt{M}_*/
\dt{M}_{SS}$ for the same $\alpha_\nu$ values.
     A summary of results for these runs is shown in Table \ref{table:wedge}.
 The most likely physical cases are those with $\Delta \ll 1$.  The
 case $\Delta=0.5$ is of interest because it allows comparison with the analytic solution of Lovelace and Chou (1996).

    For $\Delta >0$, a shear layer forms between  the co- and counter-rotating components.    Due to the viscosity, the angular momentum of the counter-rotating matter is lost  to the co-rotating matter.  In effect, 
the angular momenta of the two components
is annihilated.     The matter with zero angular momentum then 
accretes to the star with approximately free-fall speed.
Figure 5 shows the behavior for $\Delta =0.01$.  After some time all counter-rotating matter has exchanged angular momentum with the main disc, and has accreted onto the star (left-hand panel). This excites disc oscillations and the system continues to evolve  (right-hand panel). 
  Notice that this small mass fraction of counter-rotating matter
excites large amplitude axisymmetric warping of the main disc.
  Large amplitude warping was found in the earlier axisymmetric
simulations by Kuznetsov et al. (1999).

We measure the peak mass accretion rate to the star $(\dt{M}_*)_p$ and the peak mass accretion rate normalized to the rate for a standard accretion disc $(\dt{m})_p$. 
     For $\Delta \leq 0.1$
we indicate the time $t_w$ for which all the counter-rotating matter has accreted onto the star. 
   Figure 6 shows the logarithm of the  mass accretion rate to the star as a function of
time for different $\Delta$ values.
   Figure 7 shows on a linear scale the mass accretion rate as a function of time for $\Delta=0.01$.

     For $\Delta \leq 0.1$, there is an initial period of rapid accretion during which angular momentum is exchanged between the counter-rotating 
 layer and the main part of the disc. 
          The mass accretion rate peaks at a rate 
  $(\dt{M}_*)_p$ of $10^2 - 10^4$  times larger than 
  the accretion rate of a conventional disc rotating in one direction.
       After the counter rotating matter is accreted, the accretion
 rate drops down to the value  of conventional disc.
  
   The duration of the period of enhanced accretion $t_w$ does
 not depend significantly on  the amount of counter-rotating matter.
 Also, it does not depend on the viscosity of the disc. 
     This supports the conclusion that a layered counter-rotating system
     of initial radius $R$  will be transient, existing only on the free-fall time scale $t_{\rm ff}  \sim R^{3/2}/\sqrt{GM}$. 
      This is consistent with our  simulation results for a counter rotating
 torus.

As a consistency check that our results are not due to some asymmetry in our code, the $\Delta = 0.01$ run was repeated with counter rotating
layers both above and below the equatorial plane. This result is included above, indicated by an asterisk, and we see that inclusion of the second
layer simply doubles the amount of matter accreting onto the star 
but does not change the accretion time. 
The reported shear layer thicknesses are the average of the upper and lower shear layers. We note that the layers thickness grows slightly more slowly, but by $t = 15$ has reached the same thickness as with the single layer case. This is  likely due to the pressure from the opposite shear layer. This pressure is responsible in the single layer case for the rising of the shear layer. In this case, it acts to compress the opposite shear layer and slow it's growth.

The thickness of the shear layer is determined by the angular momentum transfer rate between the counter-rotating components and by the angular momentum required for that matter to begin to free fall. 
      For larger  counter-rotating mass fractions, the growth rate of the layer thickness is smaller as more angular momentum is required to slow the layer down. 
      Increasing the viscosity increases the rate of angular momentum transfer and hence the thickness of the layer.   
              Note that the case $\Delta = 0.5$ can be compared with the
analytic self-similar solution of Lovelace and Chou (1996)
which predicts a shear layer thickness $\Delta Z \propto \sqrt{\alpha_\nu} $.
        Figure 8 shows the radial and angular velocity profiles as a function of $Z$ at different radii at $t=15$ for $\Delta=0.5$.
              The shear layer thickness $\Delta Z(R)$ is taken to
be  the full width at half maximum of the radial velocity profiles. 
     We find $\Delta Z \propto \alpha_\nu^{0.59 \pm 0.03}$ as shown
in Figure 9.
  There is reasonable agreement between the theory and
simulations even though the theoretical model is 
independent of time.
      A similar analysis for the $\Delta = 0.01$ case gives $\Delta Z \propto \alpha_\nu^{0.42 \pm 0.08}$.  This suggests that the steady
state behaviour of the shear layer is largely insensitive to the mass
fraction of the counter-rotating layer.


    For $\Delta=0.5$,  Lovelace and Chou (1996) also predicted the ratio of mass accretion rates of the counter-rotating disc to the 
standard Shakura-Sunyaev rate and found $\dt{m}=\dt{M}_*/\dt{M}_{SS} \sim (R/h)^2 \alpha_\nu^{-1/2}$. 
  Our  mass accretion rates for counter rotating discs are time dependent, so that we tried accretion rates at various times and found the results to be largely insensitive to the  time chosen. 
    Figure 10 shows the data and
the power law fit  which gives
 $\dt{m} \propto \alpha_\nu^{-1.33}$.
     This dependence is 
insensitive to whether we fit to the peak mass accretion rate
or  the time averaged accretion rate.    The dependence on
$\alpha_\nu$ is much stronger than predicted by Lovelace and Chou (1996).

   The viscous stress term $\tau_{ZR}\propto - \partial v_R/\partial Z$ in equation (5) is found to have an essential role in the counter rotating disc flows.  
    Without this term the shear layer thickness in some cases is
artificially thin and set by the grid resolution.

\begin{figure}
                \centering
                \includegraphics[width=.45\textwidth]{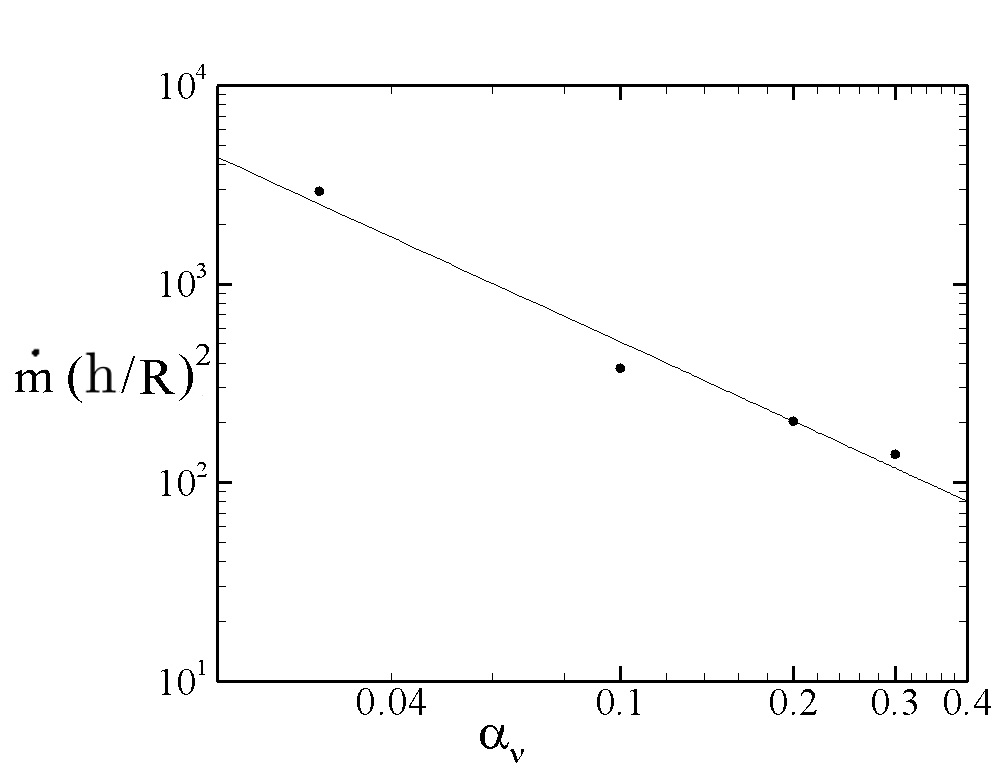}
        \caption{Mass accretion rate to the star as a function of 
  the viscosity coefficient $\alpha_{\nu}$ for $\Delta=0.5$.  
  The fit yields a power law dependence of $\dot{m} \propto (R/h)^2 \alpha_{\nu} ^ {-1.33 \pm 0.03}$, in disagreement with the self-similar solution $\dot{m} \propto (R/h)^2 \alpha_{\nu} ^ {-1/2}$ of Lovelace and Chou  (1996).
}
\label{fig:mdot_t=20_scaled}
\end{figure}

\begin{figure*}
                \centering
                \includegraphics[width=0.9\textwidth]{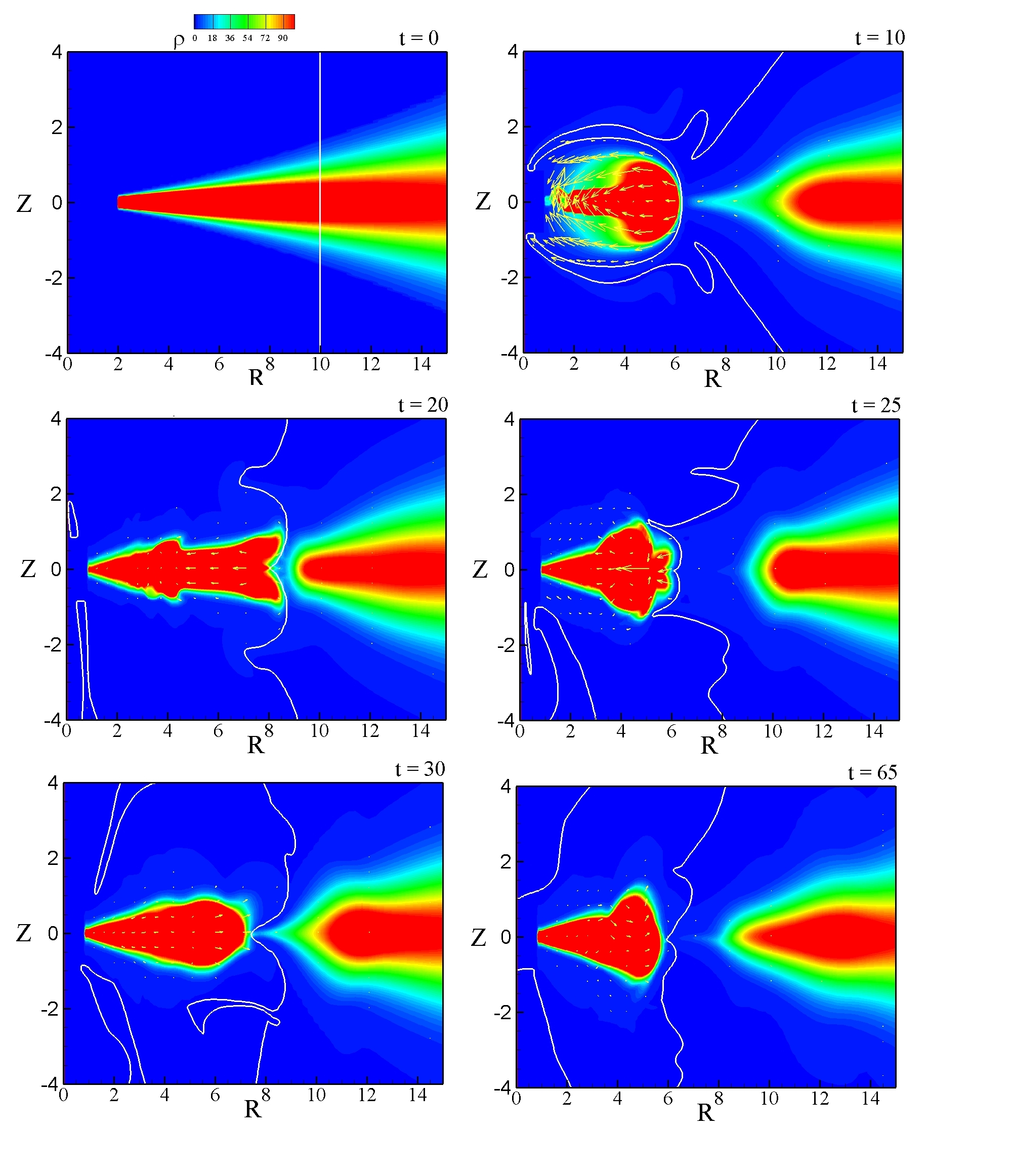}
        \caption{Plot of density $\rho$ (color) and angular velocity $\Omega = 0$ countour (white line) for $t = 0,10,20,25,30,65$. 
        The initial radius of the boundary between the co-
 and counter-rotating components $R_c = 10$.  A gap rapidly opens in the
disc layer due to the mixing of the two components.  This in turn 
excites a radial, axisymmetric breathing mode. 
Subsequently,  the inner disc oscillates quasi-periodically.}
\label{fig:anulus_summary}
\end{figure*}

\begin{figure}
                \centering
                \includegraphics[width=0.45\textwidth]{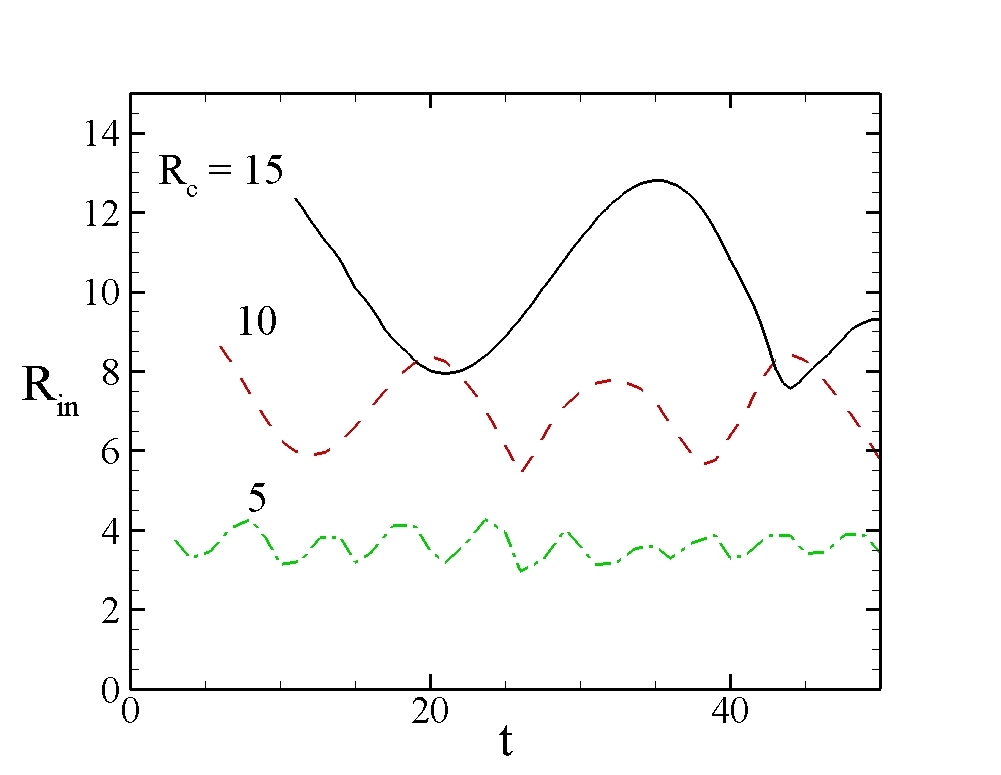}
        \caption{Outer radius of the inner disc as a function of time for initial counter-rotating boundary at $R_c = 5,10,15$. The frequency of the  oscillations is approximately the radial epicyclic frequency at the
  average value of $R_{\rm in}$.}
\label{fig:oscillations}
\end{figure}

\begin{figure}
                \centering
                \includegraphics[width=0.45\textwidth]{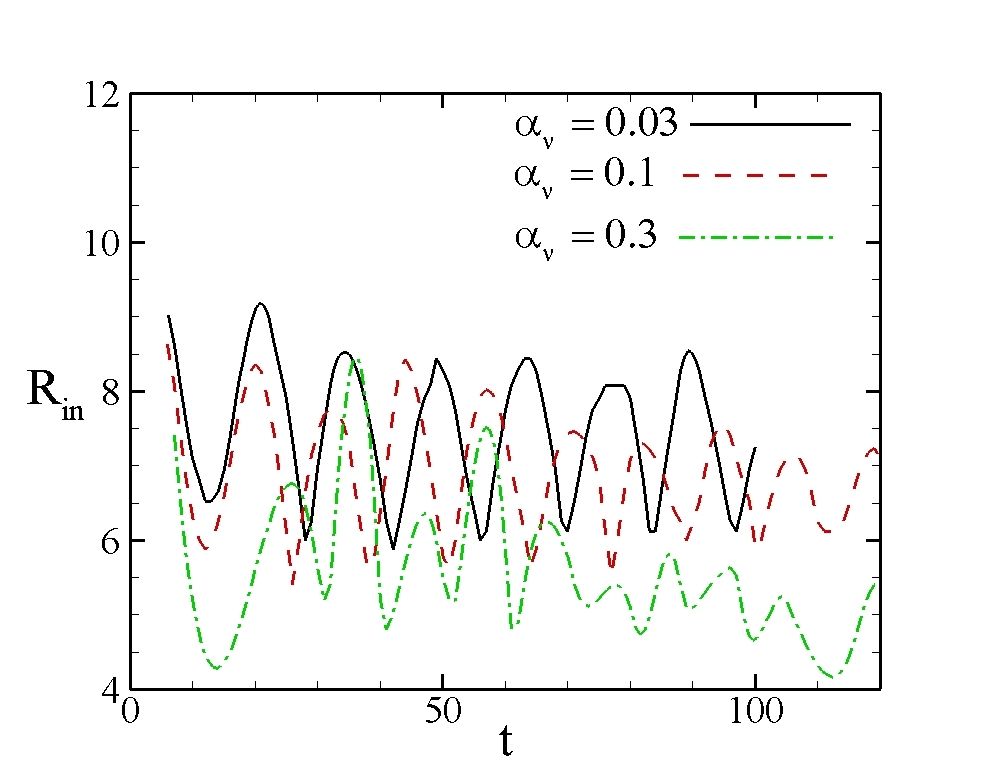}
        \caption{Outer radius of the inner disc as a function of time for initial counter-rotating boundary at $R_c=10$ and
viscosity coefficients $\alpha_{\nu} = 0.03, 0.1, 0.3$. The frequency is 
only weakly dependent on the disc viscosity. Increasing disc viscosity causes the minima of the oscillations to decrease more rapidly. The amplitude of the oscillations also decreases.}
\label{fig:oscillations_alpha}
\end{figure}

\subsection{Radially
 Separated Components}
\label{sec:anulus}

  an inner corotating Keplerian disc and an outer 
counter-rotating Keplerian disc relevant to the ``Evil Eye''
galaxy NGC 4826 (Braun et al. 1994;
Rubin 1994b) although the galaxy's  rotation curve is not Keplerian.
       The top left-hand panel of Figure 11 shows the initial disc
which has a jump discontinuity in the rotation
direction at $R_c = 10$.
       Although the disc is in mechanical equilibrium, a gap 
 rapidly  forms between the co- and counter-rotating
 components (top right-hand panel).   
     Owing to the viscosity there is mixing of the two
oppositely rotating components.   The angular momenta of
the components of the mixed gas is annihilated so that the
centrifugal force vanishes.   This gas then falls inward towards
the disc's center.   As a result the outer radius of the inner
co-rotating disc begins to oscillate  as matter piles up on the inner edge of the gap (bottom left-hand panel). 
These oscillations damp and the gap begins to fill as matter from 
the counter-rotating disc moves inward due to the viscosity
 (bottom right-hand panel).

      Figure 12 shows the oscillations of the outer radius of the 
inner disc  $R_{\rm in}(t)$ for different values of $R_c$.  
    The angular frequency of the oscillations $\omega_R$ can
 be compared with the radial epicyclic frequency
 (which for a Keplerian disc is the same as $\Omega_K$)
 evaluated at average value of $R_{\rm in}(t)$, $\overline{\kappa}$.
     For $R_c=5, 10, 15$, we find $\omega_R/\overline{\kappa} =
 1.37, 1.56.$ and $1.91$, respectively.
  
   In addition to the oscillations of $R_{\rm in}$, the inner and outer edges of the gap slowly move inward due to viscosity. 
       Figure 13 shows this inward motion for different values of
  the viscosity coefficient $\alpha_\nu$.   
       The inward motion is enhanced at larger viscosities, but the
  frequency of the radial oscillations does not change
significantly.

We find that the average mass accretion rates to the star are $\sim 25$ times larger than that of a conventional disc rotating in one direction  with the same viscosity.    
     This can be understood qualitatively as follows. 
The contact of the oppositely rotating components leads to
 the annihilation of the angular momenta of the
mixed gas.
   This non-rotating gas then tends to free-fall towards
the star by moving over the surface of the inner disc.   As
it moves inward it excites the radial oscillation of the inner
disc shown in Figures 12 and 13.
       The increased accretion rate is
largely independent of the initial position of the counter-rotating boundary. Moving the boundary from $R_c = 10$ to $R_c =15$ increases the mass accretion rate by a factor of $1.3$.

\section{Conclusion}

   Observations of galaxies and models of accreting systems point to
the  occurrence of counter-rotating discs.   Such discs
may arise from the accretion of a counter-rotating gas onto the
surface of an existing co-rotating disc or from the addition of 
counter-rotating gas at the outer radius of the existing disc.
       Axisymmetric 2.5D hydrodynamic simulations have been
 used to study two main cases -- one where the oppositely
 rotating components are vertically separated and the other where
 they are radially separated. The simulations incorporate an
 isotropic $\alpha-$viscosity including all of the terms in
 the viscous stress tensor including the important term $\tau_{RZ}$. 
      Near the interface, the two components mix and
their angular momenta  annihilate.   This mixed gas has
no centrifugal support and consequently  tends to free-fall
toward the star.
  
       For vertically separated components we investigated a range of
mass fractions $\Delta$ in the counter-rotating component.   
    The simulations indicate that the counter-rotating
matter extending out to a radius $R$ accretes to the star on a time-scale
about three times the free-fall time from this radius independent of 
$\Delta$.     The factor of three is likely due to the time required for
the mixing of the two components.
     The accretion rate to the star can be enhanced by factors $10^2$ to $10^4$  over
the rate for a disc rotating in one direction with the same viscosity. 
      Such a  burst of accretion could  lead to a comparably enhanced burst
 of a magnetically driven jet (e.g., Lii et al. 2012).
     Future investigation of counter-rotating systems including a large scale magnetic field may confirm this.   
     
     The presence of a counter rotating  layer is observed
 to excite a large amplitude axisymmetric  warp in the disc similar to
 the warping observed in the earlier study by Kuznetsov et al. (1999).
   For the case where the two components have the same mass the 
simulation results are compared with the analytic self-similar
solution of Lovelace and Chou (1996).   The dependence of
the shear layer thickness on the viscosity agrees approximately
with this theory but not with predicted mass accretion rate dependence
on viscosity.

    For the case of radially separated components we observed a more
modest increase by a factor $\sim 25$ in the accretion rate over that
of a conventional disc.  The mixed gas at the interface tends to fall
inward and this is observed to  excite radial oscillations of the inner disc
with  a frequency of the order of the radial epicyclic frequency.
  This enhanced accretion is significant for the King and Pringle (2006, 2007) picture of massive black hole accretion discs  built up from a sequence of clouds with random angular momenta which  gives a disc with
rotation direction alternating as a function of radius.

     Important future work will be full 3D hydrodynamic simulations 
including the heating, cooling, and radiative transfer of the gas
of counter-rotating discs.   This will account for the supersonic
Kelvin-Helmholtz instability, which is a 3D instability, as well as
the strong shocks which it induces (see Quach et al. 2014).  
   The time-scale of mixing of the two components and
their loss of angular momentum is expected to
be shortened compared with the present results.

\section*{Acknowledgments}

     This work  was supported in part by NASA grants NNX11AF33G
and NSF grant AST-1211318.


\label{lastpage}


\begin{thebibliography}{99}

\bibitem{} Alig, C., Schartmann, M., Burkert, A., \& Dolag, K. 2013,
ApJ, 771, 119

\bibitem{} Braun, R., Walterbos, R.A.M., Kennicutt, R.C., Tacconi, L.J., 1994, ApJ, 420, 558

\bibitem{} Chakrabarty, D., Bildsten, L.,  Finger, M. H.,  Grunsfeld, J.  M.,  Koh, D. T.,  Nelson, R. W., Prince, T. A.,  Vaughan, B. A., \& Wilson, R. B. 1997, ApJ, 481, L101

\bibitem{} Choudhury, S.,R., Lovelace, R.V.E, 1983, ApJ, 283, 331

\bibitem{} Ciri, R., Bettoni, \& Galletta, G. 1995, Nature, 375, 661

\bibitem{} Comins, N.F., Lovelace, R.V.E., Zeltwanger, T., \&
Shorey, P. 1997, ApJ, 484, L33

\bibitem{} Corsini, E.M. 2014, in {\it Counter-Rotation in
Disk Galaxies}, ASP Conference Series, Vol. 486, Eds. E. Iodice
\& E. M. Corsini (ASP: San Francisco), p. 51

\bibitem{} Galetta, G. 1996, in {\it  Barred Galaxies}, 
IAU Colloq. 157, Eds.  R. Buta, D. Crocker, \& B. Elmegreen, ASP Conference Series, 91, 11

\bibitem{} Gulati, M., Tarun, D.S., \& Sridhar, S. 2012, MNRAS, 424, 348 

\bibitem{} King, A.R., \& Pringle, J.E. 2006, MNRAS, 373, L90

\bibitem{} King, A.R., \& Pringle, J.E. 2007, MNRAS, 377, L25

\bibitem{} Kuznetsov, O. A., Lovelace, R. V. E., Romanova, M. M., \& Chechetkin, V. M. 1999, ApJ, 514, 691



\bibitem{} Li, L., \& Narayan, R., 2004, ApJ, 601, 414

\bibitem{} Lii, P., Romanova, M.M., \& Lovelace, R.V.E. 2012, MNRAS, 420, 202

\bibitem{} Lovelace, R.V.E., \& Chou, T., 1996, ApJ, 468, L25

\bibitem{} Lovelace, R.V.E., Jore, K.P., \& Haynes, M.P. 1997,
ApJ, 475, 83

\bibitem{} Nelson, R. W.,  Bildsten, L., Chakrabarty, D., Finger, M. H., Koh, D. T., Prince, T.A.,  Rubin, B. C.,  Scott, D. M., Vaughan, B. A., \& Wilson, R, B. 1997, ApJ, 488,  L117

\bibitem{} Nixon, C.J., King, A.R., Price, D.J., 2012, MNRAS, 422, 2547

\bibitem{} Quach, D., Dyda, S., \& Lovelace, R.V.E. 2014, MNRAS,
submitted

\bibitem{} Ray, T.P. 1981, MNRAS, 196, 195

\bibitem{} Ray, T.P. 1982, MNRAS, 198, 617

\bibitem{} Remijan, A.J., \& Hollis, J.M. 2006, ApJ, 640, 842

\bibitem{} Rubin, V.C., Graham, J.A., Kenney, J.D.P. 1992, ApJ, 394, L9

\bibitem{} Rubin, 1994, AJ, 107, 173

\bibitem{} Rubin, 1994, AJ, 108, 456

\bibitem{} Lii, P., Romanova, M.M., \& Lovelace, R.V.E. 2012,
MNRAS, 420, 2020

\bibitem{} Sage, L.J., \& Galletta, G., 1994, ApJ, 108, 1633S.

\bibitem{SS1973} Shakura, N.I., \& Sunyaev, R.A. 1973, A\&A, 24, 337

\bibitem{} Tang, Y.W., Guilloteau, S., Pietu, V., Dutrey, A., Ohashi, N., \& Ho, P.T.P., 2012, A \& A, 547, A84

\bibitem{} Vorobyov, E.I., Lin, D.N.C., \& Guedel, M. 2014,
A\&A, in press




\end{thebibliography}
\end{document}